\magnification=1090 \vsize=25truecm \hsize=16truecm \baselineskip=0.6truecm
\parindent=1truecm \nopagenumbers \font\scap=cmcsc10 \hfuzz=0.8truecm
\def\xup{\overline x}
\def\xdo{\underline x}
\def\ydo{\underline y}
\def\Fup{\overline F}
\def\Fdo{\underline F}

\def\Gup{\overline G}
\def\Gdo{\underline G}

\def\Hup{\overline H}
\def\Hdo{\underline H}

\def\Hdd{\underline{\Hdo}}
\def\Kup{\overline K}
\def\Kdo{\underline K}

\def\Kdd{\underline{\Kdo}}
\def\Mup{\overline M}
\def\Mdo{\underline M}
\def\Mdd{\underline{\Mdo}}
\def\Nup{\overline N}
\def\Ndo{\underline N}
\def\Ndd{\underline{\Ndo}}

\def\Pdo{\underline P}

\def\Qdo{\underline Q}

\def\hdot{\!\cdot\!}

\def\dPIVvt{{3.16}}
\def\dPIVbfa{{3.17a}}
\def\dPIVbfb{{3.17b}}
\def\dPIVbfc{{3.17c}}
\def\dPIVbfd{{3.17d}}
\def\dPIVbfe{{3.17e}}
\def\dPIVbff{{3.17f}}
\def\cPIVbfa{{3.18a}}
\def\cPIVbfb{{3.18b}}
\def\cPIVbfc{{3.18c}}
\def\cPIVbfd{{3.18d}}
\def\cPIVbfe{{3.18e}}
\def\cPIVbff{{3.18f}}
\def\cPIV{{3.19}}
\def\dPIVbfg{{3.20}}
\def\cPIVbfg{{3.21}}

\def\dPV{{3.22}}
\def\dPVvt{{3.23}}
\def\dPVbfa{{3.24a}}
\def\dPVbfb{{3.24b}}
\def\dPVbfc{{3.24c}}
\def\dPVbfd{{3.24d}}
\def\dPVbfe{{3.24e}}
\def\dPVbff{{3.24f}}
\def\cPVbfa{{3.25a}}
\def\cPVbfb{{3.25b}}
\def\cPVbfc{{3.25c}}
\def\cPVbfd{{3.25d}}
\def\cPVbfe{{3.25e}}
\def\cPVbff{{3.25f}}
\def\cPV{{3.26}}

\def\dPVIa{{4.1}}
\def\dPVIb{{4.2}}


\null \bigskip  \centerline{\bf FROM DISCRETE TO CONTINUOUS PAINLEV\'E
EQUATIONS:}

\bigskip\centerline{\bf A BILINEAR APPROACH}
\vskip 2truecm

\centerline{\scap Y. Ohta}
\centerline{\sl Department of Applied Mathematics}
\centerline{\sl Faculty of Engineering}
\centerline{\sl Hiroshima University}
\centerline{\sl 1-4-1 Kagamiyama, Higashi-Hiroshima}
\centerline{\sl 739, Japan}
\bigskip
\centerline{\scap A. Ramani}
\centerline{\sl CPT, Ecole Polytechnique}
\centerline{\sl CNRS, UPR 14}
\centerline{\sl 91128 Palaiseau, France}
\bigskip
\centerline{\scap B. Grammaticos}
\centerline{\sl LPN, Universit\'e Paris VII}
\centerline{\sl Tour 24-14, 5${}^{\grave eme}$\'etage}
\centerline{\sl 75251 Paris, France}
\bigskip
\centerline{\scap K.M. Tamizhmani}
\centerline{\sl Departement of Mathematics}
\centerline{\sl Pondicherry University}
\centerline{\sl Kalapet, Pondicherry, 605104 India}

\vskip 2truecm \noindent Abstract \smallskip
\noindent  We present the bilinear forms of the (continuous) Painlev\'e
equations obtained from
the continuous limit of the analogous expresssions for the discrete ones.
The advantage of this
method is that it leads to very symmetrical results. A new and interesting
result is the
bilinearization of the P$_{\rm VI}$ equation, something that was missing
till now.

\vfill\eject

\footline={\hfill\folio} \pageno=2

{\scap 1. Introduction}
\medskip
\noindent The study of discrete Painlev\'e equations (d-P's) has greatly
profited from the
parallel that exists between these discrete systems and their continuous
counterparts. Once it
was established that most properties of the continuous Painlev\'e equations
could be transposed
{\sl mutatis mutandis} to the discrete setting, the problem was greatly
simplified. (It is much
simpler to find something when you are convinced that it exists). Thus the
vast majority of the
existing results consists in establishing the discrete analog of something
that is well known
for the continuous Painlev\'e equations. The reverse process, starting from
some discrete
property and obtaining a new result for the continuous case, had not, to
our knowledge,
materialized yet. In this work we shall present such an approach for the
bilinearization of the
continuous Painlev\'e equations. In two recent work of ours we have, in
fact, presented the
discrete bilinear form for all d-P's [1,2]. (Let us point out here that in
some cases the
bilinear form was not sufficient and we had to resort to trilinear forms,
while higher
multilinear ones are also mandatory in some cases).

Our approach was based on the observation that the bilinear structure is
related in a certain way
to the singularity structure of the mapping. More specifically, the number
of different
singularity patterns was an indication as to the number of $\tau$-functions
necessary for the
bilinearization. Moreover, the precise structure of the singularities was
dictacting the
expression of the nonlinear variable in terms of the $\tau$-functions. As a
result of this
approach, we were able to obtain simple, highly symmetrical bilinear
expressions for the discrete
Painlev\'e equations. In what follows, we shall use these expressions as a
starting point and, by
implementing the appropriate limit, obtain the bilinear forms for the
continuous Painlev\'e
equations [3].

In their work [4], Hietarinta and Kruskal have presented such a study,
which is close in
spirit to ours since it was based on the consideration of the singularities
of the  continuous
Painlev\'e equations. However, since the process of splitting a multilinear
equation so as to
reduce it to bilinear ones is non-systematic, in many instances, the final
expressions of [4]
were not very symmetric. This is avoided if one proceeds through the
discrete case. As a matter
of fact, the discrete setting introduces so many constraints that one is
left with very little
freedom as to the possible form of the bilinear equation. Another important
point is that in [4]
no bilinear form for P$_{\rm VI}$ could be obtained. This is remedied here.
Starting from the
discrete bilinear expression of $q$-P$_{\rm VI}$ [2], we were able to
obtain its continuous
counterpart.

A nice feature of this approach is that one does not have to worry about
the alternate forms of
d-P's, i.e. the existence of many different expressions for d-P$_{\rm I}$,
d-P$_{\rm II}$ and so
on. In the continuous case they go over to the same continuous Painlev\'e
equation. Thus
one can start from the version of the d-P with the most convenient bilinear
form and work out
its continuous limit.

\bigskip {\scap 2. The case of the discrete P$_{\rm I}$ equation}
\medskip
\noindent The d-P$_{\rm I}$ equation has a particular status in the sense
that it is the only
equation involving only one $\tau$-function. This is related to the fact
that d-P$_{\rm I}$ has
only one singularity pattern. However, when several forms of d-P$_{\rm I}$
were studied, it
turned out that these equations could not all be bilinearized. In some
cases the resulting form
was a trilinear one. In particular for the d-P$_{\rm I}$:
$$\xup+\xdo={z\over x}+{a\over x^2}\eqno(2.1)$$ where $\xup=x(n+1)$,
$\xdo=x(n-1)$ and $z=\alpha
n+\beta$, we put:
$$x={\overline {F} \underline {F} \over F^2}\eqno(2.2)$$ The latter
suggested by the singularity
structure
$\{0,\infty^2,0\}$ where $\infty^2$ is a shorthand  notation for  a
singularity that behaves as
$\epsilon^{-2}$ when the `0' corresponds just to $\epsilon$ as $\epsilon\to
0$. This results to
the form:
$$\overline {\Fup}\Fdo^2+ \underline {\Fdo}\Fup^2=z\Fup F\Fdo+aF^3\eqno
(2.3)$$ The continuous
limit of this equation is obtained through $z=6+\epsilon^4\zeta/2$, $a=-4$ while
$x=1+\epsilon^2w/2$ leads to $w=2(\log F)_{\zeta\zeta}$. We point out here
that the continuous
limit of $F$ is just $F$ considered now as a function of $\zeta$ (rather
than $n$). Thus while
(2.1) reduces to
$$w''+3w^2=\zeta\eqno(2.4)$$ the bilinear equation for $F$ becomes:
$$F(D^4_{\zeta}F\hdot F)=\zeta F^3\eqno(2.5)$$ or, after a trivial division
by $F$, to a bilinear
equation. This is in fact the bilinear equation of  P$_{\rm I}$ obtained by
Hietarinta and
Kruskal. Another form of d-P$_{\rm I}$, the so-called `standard' form, is:
$$\xup+x+\xdo={z\over x}+a\eqno(2.6)$$ and has a slightly different
trilinearization. Indeed the
substitution
$$x={\overline {\Fup} \underline {F} \over \Fup F}\eqno(2.7)$$ suggested by
the singularity
structure $\{0,\infty,\infty,0\}$ is not applied to (2.6) but rather to its
discrete derivative
leading to:
$$\overline{\overline {\Fup}}\Fdo\,\underline{\Fdo}-\overline
{\Fup}\,\Fup\underline{\underline
{\Fdo}}=z\Fup^2\underline {\Fdo}-\underline z\overline {\Fup}\Fdo^2\eqno
(2.8)$$ The continuous
limit of the latter, obtained through $z=-3+\epsilon^4\zeta$ is, expectedly:
$${d\over d\zeta}\left({D^4_{\zeta}F\hdot F\over F^2}\right)=1\eqno(2.9)$$
With $a=6$ and
$x=1+\epsilon^2 w$, so that again $w=2(\log F)_{\zeta\zeta}$, equation(2.9)
gives upon
integration in
$\zeta$ the same bilinear form of P$_{\rm I}$ as (2.5).

Finally there are also multiplicative forms of d-P$_{\rm I}$ [5], the
simplest of which is:
$$\xup\xdo={z\over x}+{a\over x^2}\eqno(2.10)$$ where in this case
$z=\mu\lambda^n$. The
singularity structure $\{0,\infty^2,0\}$ suggests again
$x={\Fup \underline {F} / F^2}$, whereupon (2.10) becomes directly the bilinear
equation
$$\overline {\Fup}\underline{\Fdo}=z\Fup\Fdo+aF^2\eqno(2.11)$$ Its
continuous limit corresponds
to $\lambda=1+\epsilon^5/8$, $\mu=4$, $a=-3$, and reads again:
$$D^4_{\zeta}F\hdot F=\zeta F^2\eqno(2.12)$$ with $\zeta=n\epsilon$. Again,
we recover (2.4)
with $x=1+\epsilon^2 w/2$. In fact, all the forms of P$_{\rm I}$ analyzed
in [1] lead to the
same continuous bilinear form, which coincides with the one obtained in
[4]. While no new result
is obtained for this simplest d-P, this will no more be the case for the
`higher' ones.

\bigskip {\scap 3. Bilinear forms for the Painlev\'e equations II to V}
\medskip
\noindent In [1] we have examined the discrete Painlev\'e equations II to V
and obtained their
bilinear forms. We have shown that the number of $\tau$-functions  needed
for the
bilinearization was as follows. For d-P$_{\rm II}$, 2 $\tau$-functions. For
d-P$_{\rm III}$, 2
or 4 $\tau$-functions, the latter choice resulting in a more symmetrical
form. For d-P$_{\rm
IV}$, 4 or 6 $\tau$-functions, where again the latter choice is  more
symmetrical, while for
d-P$_{\rm V}$ 6 $\tau$-functions are needed. Based already on this counting
argument, we expect
our results to be different from those of Hietarinta and Kruskal who in [4]
have presented the
bilinearization of the Painlev\'e equations with a number of
$\tau$-functions not exceeding
three.

\noindent Here are our results. The standard form of d-P$_{\rm II}$ is:
$$\xup+\xdo={zx+a\over 1-x^2} \eqno(3.1)$$ where $z$ is linear in $n$.
Singularity structure
suggests:
$$x=-1+{\Fup\Gdo\over FG}=1-{\Fdo\Gup\over FG}\eqno(3.2)$$ hence the
bilinear equations:
$$\Fup\Gdo+\Fdo\Gup=2FG \eqno(3.3)$$
$$\overline{\Fup}\underline{\Gdo}-\underline{\Fdo}\overline{\Gup}=z(\Fup\Gdo
-\Fdo\Gup)+2aFG
\eqno(3.4)$$ The continuous limit is obtained through
$z=2+\epsilon^2\zeta,a=\epsilon^3\alpha$:
$$D^2_{\zeta}F\hdot G=0 \eqno(3.5)$$
$$(D^3_{\zeta}-{\zeta}D_{\zeta}-\alpha)F\hdot G=0 \eqno(3.6)$$ This is
compatible with the
continuous limit to P$_{\rm II}$
$$w_{\zeta \zeta}=2w^3+\zeta w+\alpha\eqno(3.7)$$ through $x=\epsilon w$,
$w=(\log
F/G)_{\zeta}$.

\noindent The standard form of d-P$_{\rm III}$ is:
$$\xup\xdo={cd(x-az)(x-bz)\over (x-c)(x-d)} \eqno(3.8)$$ with
$z=\mu\lambda^n$. We present
first the more symmetrical bilinear form involving 4
$\tau$-functions. The singularity structures suggest:
$$x=c\left(1-{\Fup\Gdo\over FG}\right)=d\left(1-{\Fdo\Gup\over
FG}\right)={HK\over FG}$$
$${1\over x}={1\over az}\left(1-{\Hup\Kdo\over HK}\right)={1\over
bz}\left(1-{\Hdo\Kup\over
HK}\right)={FG\over HK}\eqno(3.9)$$ leading to the bilinear equations:
$$(c-d)FG-c\Fup\Gdo+d\Fdo\Gup=0 \eqno(3.10a)$$
$$(a^{-1}-b^{-1})HK-a^{-1}\Hup\Kdo+b^{-1}\Hdo\Kup=0 \eqno(3.10b)$$
$$(c+d)FG-c\Fup\Gdo-d\Fdo\Gup=2HK \eqno(3.10c)$$
$$(a^{-1}+b^{-1})HK-a^{-1}\Hup\Kdo-b^{-1}\Hdo\Kup=2zFG \eqno(3.10d)$$ The
continuous limit is
obtained through $\lambda=1+\epsilon$, $a=\epsilon-a_0\epsilon^2$,
$b=-\epsilon-b_0\epsilon^2$,
$c=1/\epsilon+c_0$, $d=-1/\epsilon+d_0$ leading to:
$$D_{\zeta}F\hdot G=-HK$$
$$D_{\zeta}H\hdot K=-zFG$$
$$D_{\zeta}^2 F\hdot G=(c_0+d_0)HK\eqno(3.11)$$
$$D_{\zeta}^2H\hdot K=(a_0+b_0)zFG$$ where $z=e^{\zeta}$. In fact, since
$x=HK/FG$, it goes, at
the continuous limit, to
$x=-z\log{(F/G)}_{z}$ (and also $-(\log{(H/K)}_z)^{-1}$) and satisfies the
continuous P$_{\rm
III}$ (though not exactly under its canonical form):
$$x_{zz}={x^2_{z}\over x}-{x_{z}\over z}+{x^3\over z^2}-{c_0+d_0\over
z^2}x^2+{a_0+b_0\over
z}-{1\over x}.\eqno(3.12)$$
Instead of (3.10) there also exists a bilinear form for d-P$_{\rm III}$
involving only the two
$\tau$-functions $F$, $G$. It consists of (3.10a) and
$${cd\over c-d}(c\overline {\overline F}\underline {\underline G}
-d\overline {\overline G}\underline {\underline F})+
(a-c)(b-d)FG+c(b-d)\overline F\underline G+d(a-c)\overline G\underline
F=0\eqno(3.13)$$
Its continuous limit is:
$$D_{\zeta}^2 F\hdot G+(c_0+d_0)D_{\zeta}F\hdot G=0$$
$$D_{\zeta}^4 F\hdot G+(c_0+d_0)D^3_{\zeta}F\hdot G+2(a_0+b_0)zD_{\zeta}^2
F\hdot
G+2z^2FG=0\eqno(3.14)$$
\noindent We now proceed to d-P$_{\rm IV}$ the standard form of which is:
$$(\xup+x)(x+\xdo)={(x+a)(x-a)(x+b)(x-b)\over (x+z+c)(x+z-c)} \eqno(3.15)$$
We introduce the transformation
$$
\eqalign{
x &= a-{H\Kdo \over FG}
   = -a-{\Hdo K \over FG}
   = b\left(1-{M\Ndo \over FG}\right)
   = -b\left(1-{\Mdo N \over FG}\right) \cr
  &= -z-c\left(1-{\Fup\Gdo \over FG}\right)
   = -z+c\left(1-{\Fdo\Gup \over FG}\right)}
\eqno(\dPIVvt)
$$
Note that although $a,b$ play a symmetrical role in (3.15), equation (3.16)
treats them in
an asymmetrical way. This is done with some hindsight, in view of the
continuous limit below.
This leads to:
$$
2aFG - H\Kdo + \Hdo K = 0
\eqno(\dPIVbfa)
$$
$$
2FG - M\Ndo - \Mdo N = 0
\eqno(\dPIVbfb)
$$
$$
2FG - \Fup\Gdo - \Fdo\Gup = 0
\eqno(\dPIVbfc)
$$
$$
H\Kdo + \Hdo K = 2zFG - c(\Fup\Gdo - \Fdo\Gup)
\eqno(\dPIVbfd)
$$
$$
b(M\Ndo - \Mdo N) = 2zFG - c(\Fup\Gdo - \Fdo\Gup)
\eqno(\dPIVbfe)
$$
$$
\Hup\Kdd - \Hdd\Kup + a(\Mup\Ndd + \Mdd\Nup) + {2a \over c^2}(a^2-b^2)FG = 0
\eqno(\dPIVbff)
$$
At the continuous limit ($a=\epsilon a_1$, $b=2/\epsilon +\epsilon b_1$,
$c=1/\epsilon +\epsilon
c_1$) we have:

$$
D_z H\hdot K = 2a_1FG
\eqno(\cPIVbfa)
$$
$$
MN = FG
\eqno(\cPIVbfb)
$$
$$
D_z^2 F\hdot G = 0
\eqno(\cPIVbfc)
$$
$$
HK = zFG - D_z F\hdot G
\eqno(\cPIVbfd)
$$
$$
D_z M\hdot N = zFG - D_z F\hdot G
\eqno(\cPIVbfe)
$$
$$
D_z^3 H\hdot K + 2a_1D_z^2 M\hdot N = 8a_1(b_1-2c_1)FG
\eqno(\cPIVbff)
$$
It is straightforward to check that the continuous variable
$x=-z+(\log F/G)_z=-HK/FG=-(\log M/N)_z$ satisfies the continuous P$_{\rm IV}$:

$$
x_{zz}
= {x_z^2 \over 2x} + {3\over2}x^3 + 4zx^2 + 2z^2x - {2a_1^2 \over x}
+ 2(b_1-2c_1)x
\eqno(\cPIV)
$$
It is also easy to obtain for P$_{\rm IV}$ a bilinear expression involving
only four
$\tau$-functions. Eliminating $M,N$ between (\dPIVbfb), (\dPIVbfe) and
(\dPIVbff) we
find:
$$
\Hup\Kdd - \Hdd\Kup
+ {2a \over c^2}\Bigl\{ (-z-c+b)(-z+c-b)FG+ c(-z+c-b)\Fup\Gdo -
c(-z-c+b)\Fdo\Gup \Bigr\}= 0
\eqno(\dPIVbfg)$$
and the continuous equivalent, which amounts to eliminate $M$ and $N$
between (\cPIVbfb),
(\cPIVbfe) and (\cPIVbff):
$$
D_z^3 H\hdot K = 4a_1zD_z F\hdot G - 2a_1(z^2-4b_1+8c_1)FG
\eqno(\cPIVbfg)
$$

\noindent For d-P$_{\rm V}$:
$$
(\xup x-1)(x\xdo -1)
= {pq(x-u)(x-{1 \over \displaystyle u})(x-v)(x-{1 \over \displaystyle v}) \over
   (x-p)(x-q)}
\eqno(\dPV)
$$
we introduce:
$$
\eqalign{
x &= u+{H\Kdo \over FG}
   = {1 \over u}+{\Hdo K \over FG}
   = v+{M\Ndo \over FG}
   = {1 \over v}+{\Mdo N \over FG} \cr
  &= p\left(1-{\Fup\Gdo \over FG}\right)
   = q\left(1-{\Fdo\Gup \over FG}\right)}
\eqno(\dPVvt)
$$
and find
$$
\left(u-{1 \over u}\right)FG + H\Kdo - \Hdo K = 0
\eqno(\dPVbfa)
$$
$$
\left(v-{1 \over v}\right)FG + M\Ndo - \Mdo N = 0
\eqno(\dPVbfb)
$$
$$
(p-q)FG - p\Fup\Gdo - q\Fdo\Gup = 0
\eqno(\dPVbfc)
$$
$$
\left(u+{1 \over u}\right)FG + H\Kdo + \Hdo K = (p+q)FG - p\Fup\Gdo - q\Fdo\Gup
\eqno(\dPVbfd)
$$
$$
\left(v+{1 \over v}\right)FG + M\Ndo + \Mdo N = (p+q)FG - p\Fup\Gdo - q\Fdo\Gup
\eqno(\dPVbfe)
$$
$$
{1 \over u-{1 \over \displaystyle u}}\left({1 \over u}\Hup\Kdd -
u\Hdd\Kup\right)
- {1 \over v-{1 \over \displaystyle v}}\left({1 \over v}\Mup\Ndd -
v\Mdd\Nup\right)
= - \left(u+{1 \over u}-v-{1 \over v}\right)FG
\eqno(\dPVbff)
$$
For the continuous limit we take
$u=1+\epsilon u_1$, $v=-1-\epsilon v_1$, $p=(1/\epsilon +p_0)/z$,
$q=(-1/\epsilon +p_0)/z$,
and find:
$$
2u_1FG - D_\zeta H\hdot K = 0
\eqno(\cPVbfa)
$$
$$
2v_1FG + D_\zeta M\hdot N = 0
\eqno(\cPVbfb)
$$
$$
D_\zeta^2 F\hdot G - 2p_0D_\zeta F\hdot G = 0
\eqno(\cPVbfc)
$$
$$
z(FG + HK) - D_\zeta F\hdot G = 0
\eqno(\cPVbfd)
$$
$$
z(FG - MN) + D_\zeta F\hdot G = 0
\eqno(\cPVbfe)
$$
$$
{1 \over u_1}D_\zeta^3 H\hdot K - {1 \over v_1}D_\zeta^3 M\hdot N
+ 2D_\zeta^2 H\hdot K - 2D_\zeta^2 M\hdot N = 0
\eqno(\cPVbff)
$$
where
$z=e^\zeta$. At the continuous limit we have $x=(\log
F/G)_z=1+HK/FG=-1+MN/FG$. Putting
$x=(1+w)/(1-w)$ we find that $w$ obeys the continuous P$_{\rm V}$ equation
in the form:
$$
w_{zz}
= \left({1 \over 2w} + {1 \over w-1}\right)w_z^2 - {1 \over z}w_z
+ {(w-1)^2 \over 2z^2}\left(v_1^2w - {u_1^2 \over w}\right)
- {4p_0w \over z} - {2w(w+1) \over w-1}
\eqno(\cPV)
$$

An interesting novel feature is that some of the continuous bilinear
equations are
non-differential, e.g. (\cPIVbfb) or the difference of (\cPVbfd) and
(\cPVbfe), relating 4 or 6
$\tau$-functions. This may be an explanation as to why our bilinear
expressions were not
obtained through the direct search of Hietarinta and Kruskal although these
authors also used
systematically the singularity structure argument (but directly on the
continuous equations).

\bigskip {\scap 4. Bilinear form of P$_{\rm VI}$}
\medskip
\noindent This is the most interesting result of this paper since it
provides the bilinearization
of P$_{\rm VI}$ that was unknown up to now.  The main factor for this
progress was the recent
derivation of a discrete form of P$_{\rm VI}$ by Jimbo and Sakai [6]. The
$q$-P$_{\rm VI}$
equation is written in form of a system:

$$
\xup x = {(y-\alpha \tilde z)(y-\beta \tilde z) \over (y-\gamma)(y-{1 \over
\displaystyle
\gamma})}
\eqno(\dPVIa)
$$
$$
y\ydo = {(x-az)(x-bz) \over (x-c)(x-{1 \over \displaystyle c})}
\eqno(\dPVIb)
$$
where $z=\mu\lambda^n$, $\tilde z=z\sqrt{\lambda}$ and we have the constraint
$ab=\alpha\beta$. The $\tau$-functions are introduced through:

$$
x = c\left(1+(1-z)^{1/2}{M\Ndo\over FG}\right)
  = {1 \over c}\left(1+(1-z)^{1/2}{\Mdo N\over FG}\right)
  = {HK\over FG}
$$
$$
{1 \over x} = {1 \over az}\left(1-(1-z)^{1/2}{P\Qdo\over HK}\right)
            = {1 \over bz}\left(1-(1-z)^{1/2}{\Pdo Q\over HK}\right)
            = {FG\over HK}
$$
$$
y = \gamma\left(1+(1-\tilde z)^{1/2}{\Fup G \over MN}\right)
  = {1 \over \gamma}\left(1+(1-\tilde z)^{1/2}{F\Gup \over MN}\right)
  = {PQ\over MN}
\eqno(4.3)
$$
$$
{1 \over y} = {1 \over \alpha\tilde z}\left(1-(1-\tilde z)^{1/2}{\Hup K
\over PQ}\right)
            = {1 \over \beta\tilde z}\left(1-(1-\tilde z)^{1/2}{H\Kup \over
PQ}\right)
            = {MN \over PQ}
$$
leading to:
$$
2FG + (1-z)^{1/2}(M\Ndo + \Mdo N) = \left(c+{1 \over c}\right)HK
$$
$$
2HK - (1-z)^{1/2}(P\Qdo + \Pdo Q) = (a+b)zFG
$$
$$
2MN + (1-\tilde z)^{1/2}(\Fup G + F\Gup)
= \left(\gamma+{1 \over \gamma}\right)PQ
$$
$$
2PQ - (1-\tilde z)^{1/2}(\Hup K + H\Kup)
= (\alpha+\beta)\tilde z MN
$$
$$
\left(c-{1 \over c}\right)FG
+ (1-z)^{1/2}\left(cM\Ndo - {1 \over c}\Mdo N\right) = 0
\eqno(4.4)
$$
$$
\left({1 \over a}-{1 \over b}\right)HK
- (1-z)^{1/2}\left({1 \over a}P\Qdo - {1 \over b}\Pdo Q\right) = 0
$$
$$
\left(\gamma-{1 \over \gamma}\right)MN
+ (1-\tilde z)^{1/2}\left(\gamma\Fup G - {1 \over \gamma}F\Gup\right) = 0
$$
$$
\left({1 \over \alpha}-{1 \over \beta}\right)PQ
- (1-\tilde z)^{1/2}\left({1 \over \alpha}\Hup K - {1 \over
\beta}H\Kup\right) = 0
$$
We go to the continuous limit through:
$a=1+\epsilon a_1+\epsilon^2a_2$, $b=1-\epsilon a_1+\epsilon^2b_2$,
$c=1+\epsilon c_1$,
$\alpha=1+\epsilon\alpha_1+\epsilon^2\alpha_2$,
$\beta=1-\epsilon\alpha_1+\epsilon^2\beta_2$,
$\gamma=1+\epsilon\gamma_1$. The constraint $ab=\alpha\beta$ translates into
$a_2+b_2-a_1^2=\alpha_2+\beta_2-\alpha_1^2$. We then find:
$$
FG + (1-z)^{1/2}MN = HK
$$
$$
HK - (1-z)^{1/2}PQ = zFG
$$
$$
c_1FG + (1-z)^{1/2}\left(D_\zeta + c_1\right) M\hdot N = 0
$$
$$
a_1HK + (1-z)^{1/2}\left(D_\zeta - a_1\right) P\hdot Q = 0
$$
$$
\gamma_1MN + (1-z)^{1/2}\left(D_\zeta + \gamma_1\right) F\hdot G = 0
\eqno(4.5)
$$
$$
\alpha_1PQ + (1-z)^{1/2}\left(D_\zeta - \alpha_1\right) H\hdot K = 0
$$
$$
\eqalign{
&(1-z)D_\zeta^2 F\hdot G - (1-z)^{1/2}D_\zeta^2 M\hdot N +
(1-z)^{1/2}D_\zeta^2 P\hdot Q \cr
&\qquad = - (a_2+b_2)zFG - c_1^2HK + \gamma_1^2(1-z)^{1/2}PQ}
$$
$$
\eqalign{
&(1-z)D_\zeta^2 H\hdot K - z(1-z)^{1/2}D_\zeta^2 M\hdot N +
(1-z)^{1/2}D_\zeta^2 P\hdot Q \cr
&\qquad = - (a_2+b_2)zFG - c_1^2zHK - (\alpha_2+\beta_2)z(1-z)^{1/2}MN}
$$
where
$z=e^{\zeta/2}$, $x=HK/FG$ and, at the continuous limit, we have in addition
$x=1+(1-z)^{1/2}MN/FG$,
$1/x=(1-(1-z)^{1/2}PQ/HK)/z$. We obtain thus the continuous P$_{\rm VI}$:

$$
\eqalign{
x_{zz}
&= {1\over2}\left({1 \over x}+{1 \over x-1}+{1 \over x-z}\right)x_z^2
+ \left({1 \over z}+{1 \over z-1}-{1 \over z-x}\right)x_z \cr
&- {x(x-1)(x-z) \over z^2(z-1)^2}
  \left({\gamma_1^2 \over 2} - {\alpha_1^2 \over 2}{z \over x^2}
       + {c_1^2 \over 2}{z-1 \over (x-1)^2}
       + {1-a_1^2 \over 2}{z(z-1) \over (x-z)^2}\right)}
\eqno(4.6)
$$

We remark here also that there exist non-differential relations between the
$\tau$-functions and they are unavoidable for the bilinearization of
P$_{\rm VI}$.

\bigskip {\scap 5. Conclusion}
\medskip
\noindent In the previous sections, we have presented the bilinearization
of the six Painlev\'e
transcendental equations starting from the results for the discrete ones
and implementing the
appropriate continuous limit. This approach has made possible the
derivation of the bilinear form
for P$_{\rm VI}$,  a result that was obtained here for the first time.

The analogy between discrete and continuous case is also useful in a
broader scope. In our
analysis of discrete equations, it became clear that the `right' number of
$\tau$-functions is
identical to the number of different singularity patterns. The same appears
to be true in the
continuous case (with one possible exception for P$_{\rm IV}$ where the use of 6
$\tau$-functions leads to a more symmetrical result). Now that results on
discrete Painlev\'e
equations start accumulating, it would be interesting to translate them
back to the continuous
case. New results for continuous equations may thus make their appearance
and old results may be
transcribed in more symmetrical, easier to use, forms.
\bigskip
{\scap Acknowledgements}
\smallskip\noindent
The authors wish to acknowledge the financial support of the CEFIPRA (under
contract 1201-A)
that had made the present collaboration possible.
Y. Ohta acknowledges the finantial support of the Japan Ministry of
Education through the Foreign Study Program.
K.M. Tamizhmani also acknowledges the award of the fellowship from
Minist\`ere des
Affaires Etrang\`eres (France).
\noindent
\smallskip {\scap References}.
\smallskip
\item{[1]} A. Ramani, B. Grammaticos and J. Satsuma, J. Phys. A 28 (1995) 4655.
\item{[2]} M. Jimbo, H. Sakai, A. Ramani and B. Grammaticos, {\sl Bilinear
structure and
Schlesinger transforms of the $q$-P$_{\rm III}$ and $q$-P$_{\rm VI}$
equations}, in preparation.
\item{[3]} K. Okamoto, Proc. Japan Acad. A56 (1980) 264.
\item{[4]} J. Hietarinta and M.D. Kruskal, {\sl Hirota forms for the six
Painlev\'e equations
from singularity analysis}, NATO ASI B278 (New York, Plenum, 1992) p. 175.
\item{[5]} A. Ramani and B. Grammaticos, {\sl Discrete Painlev\'e
equations: coalescences, limits
and degeneracies}, preprint 1995.
\item{[6]}      M. Jimbo and H. Sakai, {\sl A $q$-analog of the sixth
Painlev\'e equation}, preprint
Kyoto-Math 95-16.

\end